\documentclass[aps,prl,twocolumn,showpacs]{revtex4}
\usepackage{graphicx,latexsym}
\pdfoutput=1

\begin{document}
\newcommand{\li}{$^6$Li $ $}
\newcommand{\na}{$^{23}$Na $ $}
\newcommand{\up}{\uparrow}
\newcommand{\down}{\downarrow}
\newcommand{\rb}{\mathrm{b}}
\newcommand{\rbb}{\mathrm{bb}}
\newcommand{\rf}{\mathrm{f}}
\newcommand{\rbf}{\mathrm{bf}}
\newcommand{\cenergy}{\mathcal{E}}

\title{Realization of a strongly interacting Bose-Fermi mixture from a two-component Fermi gas}

\author{Yong-il Shin}\email{yishin@mit.edu}
\author{Andr\'{e} Schirotzek}
\author{Christian H. Schunck}
\author{Wolfgang Ketterle}

\affiliation{Department of Physics, MIT-Harvard Center for Ultracold
Atoms, and Research Laboratory of Electronics, Massachusetts Institute of
Technology, Cambridge, Massachusetts 02139, USA}

\date{\today}

\begin{abstract}
We show the emergence of a strongly interacting Bose-Fermi mixture from a
two-component Fermi mixture with population imbalance. By analyzing
\emph{in situ} density profiles of \li atoms in the BCS-BEC crossover
regime we identify a critical interaction strength, beyond which all
minority atoms pair up with majority atoms, and form a Bose condensate.
This is the regime where the system can be effectively described as a
boson-fermion mixture. We determine the dimer-fermion and dimer-dimer
scattering lengths and beyond-mean-field contributions. Our study
realizes a Gedanken experiment of bosons immersed in a Fermi sea of one
of their constituents, revealing the composite nature of the bosons.
\end{abstract}


\maketitle

Fermions are the fundamental building blocks of matter, whereas bosons
emerge as composite particles. The simplest physical system to study the
emergence of bosonic behavior is a two-component fermion mixture, where
the composite boson is a dimer of the two different fermions.  A dramatic
manifestation of bosonic behavior is Bose-Einstein condensation,
representing the low-temperature phase of a gas of bosons.  One way to
reveal the composite nature of the bosons is to immerse such a
Bose-Einstein condensate (BEC) into a Fermi sea of one of its
constituents. The degeneracy pressure due to the Pauli exclusion
principle affects the structure of the composite boson, resulting in a
zero-temperature quantum phase transition to a normal state where
Bose-Einstein condensation is quenched.

In this paper we observe this transition experimentally. We identify the
regimes where a two-component Fermi gas can be described as binary
mixture of bosons and fermions, and where the composite nature of the
boson becomes essential. The validity of a Bose-Fermi (BF) description
requires that all minority fermions become bound as bosons and form a
BEC.  We determine the critical value of $1/k_{F\up}a$ for the onset of
superfluid behavior in the limit of large population imbalance. Here $a$
is the fermion-fermion scattering length and $k_{F\up}$ is the Fermi wave
number characterizing the depth of the majority Fermi sea. Of course, for
an equal mixture, the zero-temperature ground state is always a
superfluid in the BEC-BCS crossover.  It has been shown previously that a
crossover superfluid can be quenched by population imbalance, also called
the Chandrasekhar-Clogston (CC) limit of
superfluidity~\cite{ZSS06a,SSS08}. In this work we determine the critical
point where superfluidity can no longer be quenched by population
imbalance, i.e. the CC limit becomes 100\%.

In the limit of a BF mixture~\cite{PS06}, we observe repulsive
interactions between the fermion dimers and unpaired fermions. They are
parameterized by an effective dimer-fermion scattering length of
$a_\rbf=1.23(3)a$. This value is consistent with the exact value
$a_\rbf=1.18a$ which has been predicted over 50 years ago for the three
fermion problem~\cite{ST57}, but has never been experimentally
confirmed.  The boson-boson interactions were found to be stronger than
the mean-field prediction in agreement with the Lee-Huang-Yang
prediction~\cite{LHY57}.

\begin{figure*}
\begin{center}
\includegraphics[width=4.6in]{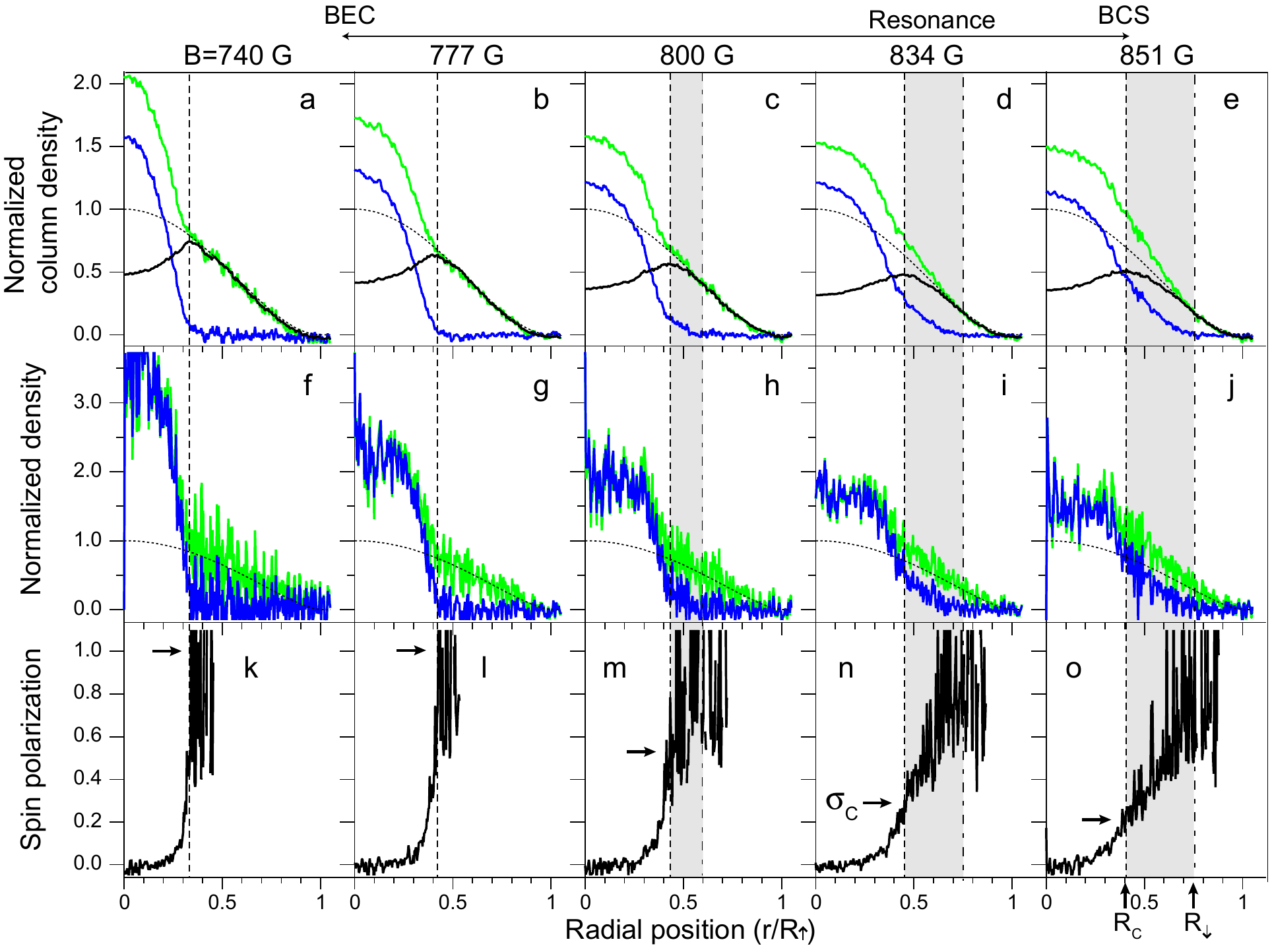}
\caption{(color online) Density profiles of imbalanced Fermi mixtures in
a harmonic trap. The top row (a-e) shows the averaged column density
profiles for various magnetic fields (green: majority, blue: minority,
black: difference). The black dotted line is a zero-temperature
Thomas-Fermi distribution fit to the majority wing profile ($r>R_\down$).
The middle row (f-j) and the bottom row (k-o) show the reconstructed
three-dimensional density distributions and the spin polarizations
obtained from the profiles in the top row. $R_\up$, $R_\down$ (dashed dot
lines), and $R_c$ (dashed lines) are the radii of the majority cloud, the
minority cloud, and the superfluid core, respectively. The critical
polarizations $\sigma_c$ at the phase boundary $r=R_c$ are indicated by
the right arrows. The values for $R_\up$ (in $\mu$m), $R_c/R_\up$, and
$R_\down/R_\up$ were respectively: for (a,f,k), 381, 0.33, 0.33; for
(b,g,l), 380, 0.33, 0.33; for (c,h,m), 362, 0.35, 0.59; for (d,i,n), 371,
0.44, 0.72; for (e,j,o), 367, 0.41, 0.76. $T/T_{F0}\lesssim 0.05$ and
$T_{F0}\approx 1.0~\mu$K (see the text for definitions).
\label{f:profiles}}
\end{center}
\end{figure*}

The system for this study is a variable spin mixture of the two lowest
hyperfine states $|\uparrow\rangle$ and $|\downarrow\rangle$ of $^6$Li
atoms (corresponding to the $|F=1/2,m_F=1/2\rangle$ and
$|F=1/2,m_F=-1/2\rangle$ states at low magnetic field) in an optical
dipole trap as described in Refs.~\cite{ZSS06a,SSS08}. A broad Feshbach
resonance, located at 834~G~\cite{BAR05}, strongly enhances the
interactions between the two spin states. The final evaporative cooling
was performed at 780~G by lowering the trap depth. Subsequently, the
magnetic-bias field $B$ is adjusted to a target value with a ramp speed
of $\leq 0.4$~G/ms, changing the interaction strength adiabatically. At
the end of the preparation, our sample was confined in an effective
three-dimensional harmonic trap with cylindrical symmetry. The axial
(radial) trap frequency was $\omega_z/2\pi= 22.8$~Hz ($\omega_r /2\pi=
140$~Hz).

The phase diagram for the fermion mixture was obtained from the analysis
of \textit{in situ} density profiles of the majority (spin $\uparrow$)
and minority (spin $\downarrow$) components. The profiles were recorded
using a phase-contrast imaging technique~\cite{SSS08}. Under the local
density approximation (LDA), low-noise column-density profiles were
obtained by averaging the optical signal along equipotenital lines (refer
to Ref.~\cite{SSS08} for a full description of the image processing). For
typical conditions, the temperature of a sample was $T/T_{F0}\lesssim
0.05$, determined from the outer region of the cloud~\cite{SSS08}, where
$T_{F0} \approx 1.0~\mu$K is the Fermi temperature of the majority
component measured as $k_B T_{F0}=m\omega_z^2 R_\up^2/2$ ($k_B$ is the
Boltzmann's constant, $m$ is the atom mass, and $R_\up$ is the axial
radius of the majority cloud).

Figure~\ref{f:profiles} displays density profiles of imbalanced Fermi
mixtures for various magnetic fields, showing how the spatial structure
of a trapped sample evolves in the crossover regime. Near resonance, as
reported in Ref.~\cite{SSS08}, three distinctive spatial regions are
identified: (I) a superfluid core, (II) an intermediate region of a
paritially-polarized normal ($N_\mathrm{pp}$) phase, and (III) a
fully-polarized, outer wing. The core radius $R_c$ was determined as the
peak position in the column density difference profile and the majority
(minority) radius $R_\up$ $(R_\down)$ was determined from the fit of the
outer region, $r>R_\up$ $(r>R_c)$ of the majority (minority) column
density profile to a zero-temperature Thomas-Fermi distribution. The
local spin polarization is defined as $\sigma(r)\equiv
(n_\up-n_\down)/(n_\up+n_\down)$, where $n_\up$ and $n_\down$ are the
local majority and minority density, respectively.

\begin{figure}
\begin{center}
\includegraphics[width=3.2in]{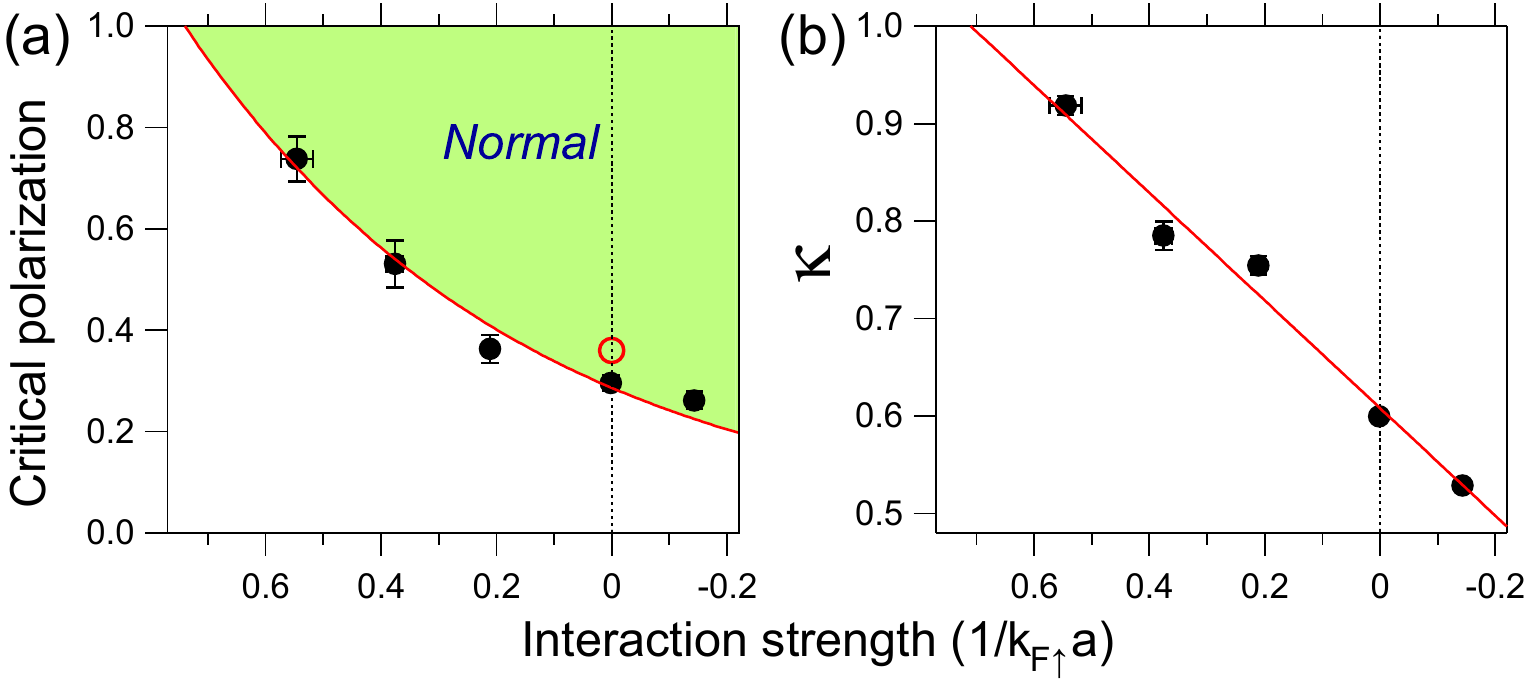}
\caption{(color online) Emergence of a Bose-Fermi mixture in the phase
diagram for a two component Fermi gas.  (a) The critical polarization
$\sigma_c$ as a function of the interaction strength $1/k_{F\up}a$ at the
phase boundary.  The open circle indicates the previously measured
critical value on resonance, $\sigma_{c0}=0.36$~\cite{SSS08}.  (b)
$\kappa=(R_\up^2-R_\down^2)/(R_\up^2-R_c^2)$. The red solid lines are (a)
an exponential fit and (b) a linear fit to the data points. $\sigma_c=1$
and $\kappa=1$ (i.e. $R_\down=R_c$) imply the absence of minority fermions
in the normal phase. Each data point consists of nine to twenty three
independent measurements and the error bars indicate only the statistical
uncertainty.\label{f:critical}}
\end{center}
\end{figure}

Further on the BEC side, the sample has a more compressed superfluid
core, a narrower intermediate normal region (gray region in
Fig.~\ref{f:profiles}) and a higher critical spin polarization at the
phase boundary $\sigma_c=\sigma(R_c)$. Eventually, when $B<780~$G, there
is no noticeable intermediate region, implying that every minority atom
pairs up with a majority atom in the superfluid core.  In
Fig.~\ref{f:critical} we determine the critical point for the
disappearance of the partially polarized normal phase in two different
ways.  Fig.~\ref{f:critical}(a) shows the phase diagram for the
$N_\mathrm{pp}$ phase in the plane of interaction strength $1/k_{F\up}a$
and spin polarization $\sigma$. An extrapolation of the critical line to
$\sigma_c=1$ yields $1/k_{F\up,c}a = 0.74(4)$. Another implication of the
absence of an $N_\mathrm{pp}$ phase is that the size of the minority
cloud $R_\down$ approaches the radius $R_c$ of the superfluid core. This
extrapolation is conveniently done using the dimensionless parameter
$\kappa=(R_\up^2-R_\down^2)/(R_\up^2-R_c^2)$~\cite{note1}, resulting in a
 value of  $1/k_{F\up,c}a = 0.71(5)$. These values are
in good agreement with recent quantum Monte-Carlo (QMC)
calculations~\cite{PG08}.

The critical point marks the onset of the emergence of a BF mixture from
the two-component Fermi system.  One may suspect that near the critical
point the equation of state of the BF mixture is complex, but we show now
that a very simple equation of state is sufficient to quantitatively
account for the observed profiles.  Due to the external trap potential,
the local chemical potential varies form zero at the edge of a cloud to a
maximum value in the center. Therefore, knowledge of the
three-dimensional density profiles of a single cloud is sufficient to
obtain the equation of state~\cite{molmer98,PS06,SSS08,Shin08,Chevy06}.

For a zero temperature mixture of bosonic dimers with density
$n_\rb=n_\down$ and mass $m_\rb=2m$ and unpaired fermions with density
$n_\rf=n_\up - n_\down$ and mass $m_\rf=m$, the energy density
$\mathcal{E}$ can be decomposed as $\mathcal{E}=\mathcal{E}_\rbb
+\mathcal{E}_\rbf+\mathcal{E}_\rf$, where $\mathcal{E}_\rbb(n_\rb)$ and
$\mathcal{E}_\rbf(n_\rb,n_\rf)$ are the boson-boson and boson-fermion
interaction energies, respectively, and $\mathcal{E}_\rf=(3/5)\alpha
n_\rf^{5/3}$ is the kinetic energy of fermions
[$\alpha=(6\pi^2)^{2/3}\hbar^2/2m_\mathrm{f}$ and $\hbar$ is the Planck's
constant]. Here we assume that the effective mass of a fermion in a
dilute mixture is same as its bare mass~\cite{BBP67}. Under the LDA, the
densities $n_\rb(r)$ and $n_\rf(r)$ in the harmonic trap should satisfy
\begin{eqnarray}
   \mu_\mathrm{f0}=&&  \alpha n_\rf^{2/3} + \frac{\partial\cenergy_\rbf}{\partial n_\rf}+ \frac{1}{2}m_\rf \omega_z^2 r^2, \label{e:muf}
   \\
   \mu_\mathrm{b0}=&&  \frac{\mathrm{d}\cenergy_\rbb}{\mathrm{d} n_\rb}  + \frac{\partial\cenergy_\rbf}{\partial n_\rb}+ \frac{1}{2} m_\rb \omega_z^2
   r^2,\label{e:mub}
\end{eqnarray}
where $\mu_\mathrm{f0}$ and $\mu_\mathrm{b0}$ are the global chemical
potentials of fermions and bosons, respectively, referenced to the trap
bottom.

\begin{figure}
\begin{center}
\includegraphics[width=3.3in]{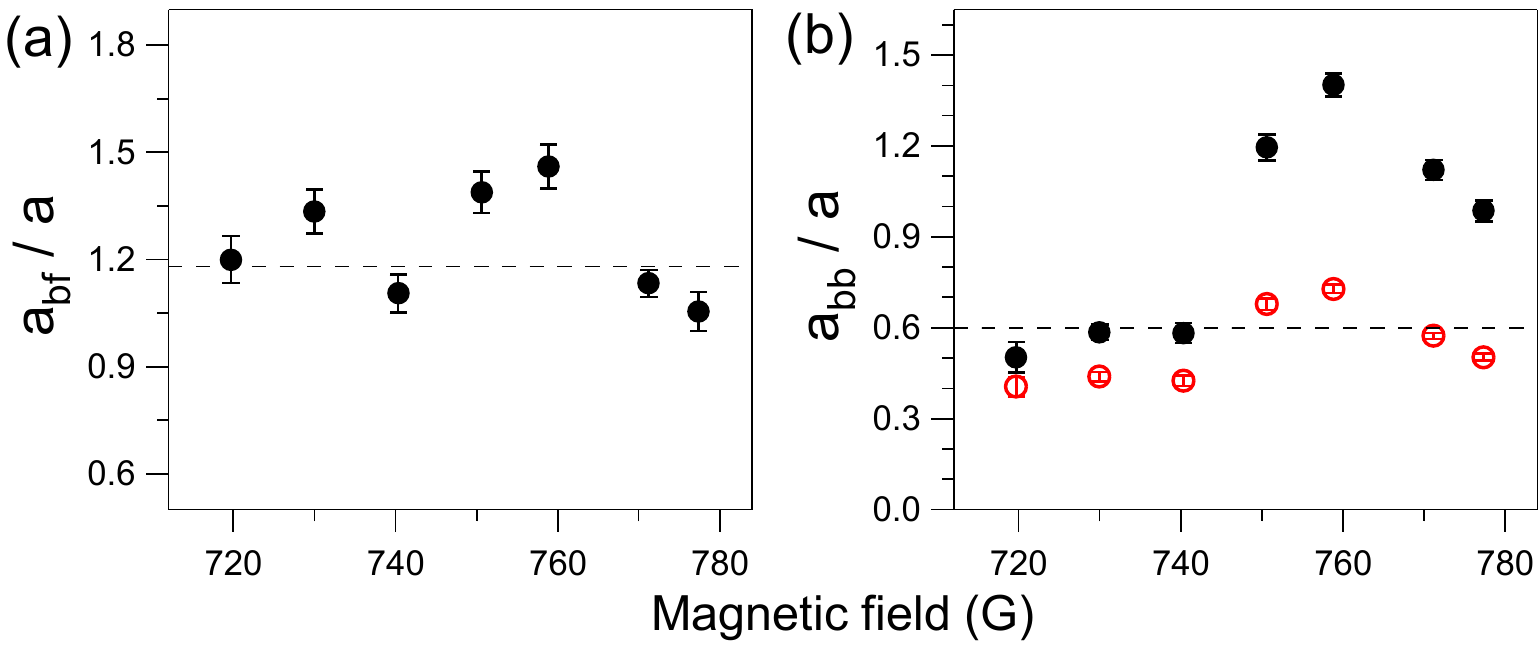}
\caption{(color online) Characterization of a strongly interacting
Bose-Fermi mixture. (a) The scattering length for dimer-fermion
interactions $a_\rbf$ and (b) for dimer-dimer interactions $a_\rbb$ in
units of the fermion-fermion scattering length $a$. Black solid circles
were determined using mean-field theory and red open circles including
the LHY correction for a strongly interacting Bose gas. The dashed lines
indicate the calculated values $a_\rbf=1.18a$~\cite{ST57} and
$a_\rbb=0.6a$~\cite{PSS04}. Each data point represents seven to seventeen
measurements and the error bars indicate only the statistical
uncertainty.\label{f:scattering}}
\end{center}
\end{figure}

For the determination of the boson-fermion scattering length $a_\rbf$, we
use a mean-field expression for the boson-fermion interaction energy
$\cenergy_{\rbf,\mathrm{M}} = (2\pi\hbar^2 /m_\mathrm{bf}) a_\rbf n_\rb
n_\rf$ with $m_\mathrm{bf}= m_\mathrm{b} m_\mathrm{f} /
(m_\mathrm{b}+m_\mathrm{f})$=(2/3)$m$. Since $\mu_\mathrm{f0}=m
\omega_z^2 R_\up^2 /2$, Eq.~(\ref{e:muf}) gives the relation,
$a_\mathrm{bf}(r)=[\mu_\mathrm{f0}(1-r^2/R_\up^2) -\alpha
n_\mathrm{f}^{2/3}]/(\frac{3 \pi \hbar^2}{m} n_\mathrm{b})$.  We obtained
a value for $a_\rbf$ by averaging $a_\rbf(r)$ over a mixed region
($r<R_c$ with $n_{\rb,\rf}>0.1 n_0$). Here, $n_0$ is the reference
density defined as $n_0=(\mu_\mathrm{f0}/\alpha)^{3/2}$.  In this
analysis, the non-interacting outer wing provides absolute density
calibration~\cite{Shin08}.

The scattering length ratio $a_\rbf/a$ turns out to be almost constant
over the whole range, 700~G$<B<$780~G where we could study BF mixtures
[Fig.~\ref{f:scattering}(a)]. For even lower magnetic fields, severe
heating occured, probably due to molecular relaxation processes. By
averaging a total of 89 measurements, we obtain $a_\rbf=1.23(3)a$, close
to the exact value $a_\rbf=1.18 a$ calculated for the three-fermion
problem~\cite{ST57}. Our finding excludes the mean-field prediction
$a_\rbf=(8/3)a$. The detailed behavior above 750~G requires further
investigation.

We now turn to the determination of the boson-boson scattering length
$a_\rbb$ which parameterizes the boson-boson mean-field energy
$\cenergy_{\rbb,\mathrm{M}}= (2\pi\hbar^2 /m_\mathrm{b}) a_\rbb n_\rb^2
$. For a given $a_\rbf$, the effective potential for bosons in the
presence of fermions is $V_\rb(r)= m \omega_z^2 r^2 + (3\pi\hbar^2/m)
a_\rbf n_\rf(r)$. Then, Eq.~(\ref{e:mub}) gives $\mu_\mathrm{b0}-
(2\pi\hbar^2/m) a_\rbb n_\mathrm{b}(r)=V_\rb(r)$. By fitting the data in
the core region ($0.1 R_\up < r< R_c$ and $n_\rb
>0.1 n_0$) to this equation with $\mu_\mathrm{b0}$ and $a_\rbb$ as two free
parameters, we obtained a value for $a_\rbb$. We used the value $a_\rbf$
determined from the corresponding profiles.

The effective mean-field values for $a_\rbb/a$ show a strong increase by
a factor of about 2, as the system approaches the critical point
[Fig.~\ref{f:scattering}(b)]. We attribute this behavior to strong
boson-boson interactions causing non-negligible quantum depletion in the
BEC. In this regime, the equation of state has to include
beyond-mean-field corrections, with the leading term given by Lee, Huang,
and Yang (LHY)~\cite{LHY57} as
\begin{equation}
   \mathcal{E}_{\mathrm{LHY}}= \frac{2\pi \hbar^2 a_\rbb n_\mathrm{b}^2}{m_\rb} \bigg[1+\frac{128}{15} \sqrt{\frac{a_\rbb^{3} n_\rb}{\pi}} \bigg].
\end{equation}
Inclusion of the LHY correction leads to smaller fitted values for
$a_\rbb/a$, which are now almost constant over the whole range of
magnetic fields with an average value of $a_\rbb=0.55(1)a$. The exact
value for weakly bound dimers is $a_\rbb=0.6a$~\cite{PSS04}. For
$k_{F,b}a=1$, the LHY correction is $0.3 \cenergy_{\rbb,\mathrm{M}}$,
i.e. a 30 \% correction to the mean field approximation.  Here,
$k_{F,\rb}=(6\pi^2 n_\rb)^{1/3}$. Recently, the LHY corrections have been
observed via the upshift of collective oscillation frequencies for a
strongly interacting BEC~\cite{ARK07}.

\begin{figure}
\begin{center}
\includegraphics[width=2.6in]{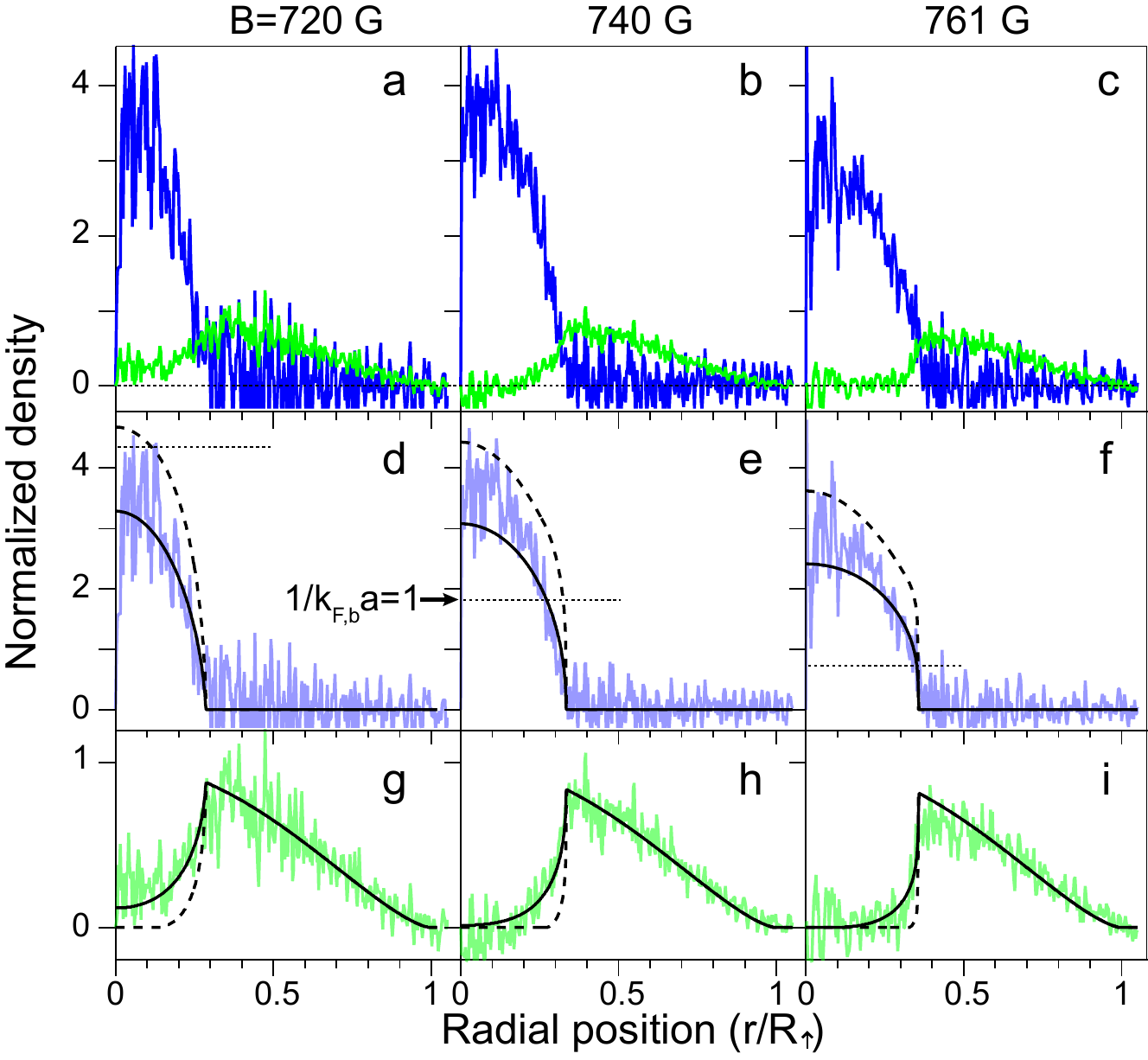}
\caption{(color online) Observed profiles of strongly interacting
Bose-Fermi mixtures compared to calculated profiles without any
adjustable parameter. (a-c) Density profiles of bosonic dimers (blue) and
unpaired excess fermions (green) for various magnetic fields. The
numerically obtained density profiles (d-f) for bosons and (g-i) for
fermions use $a_\rbb=0.6 a$ and $a_\rbf=1.18 a$ (dashed: mean-field
description, solid: including the LHY correction). The horizontal dotted
lines in (d-f) indicate the boson density corresponding to
$1/k_{F,\rb}a=1$. The values for $T_{F0}$ (in $\mu$K) and $R_c/R_\up$
were respectively: for (a), 1.15 and 0.29; for (b), 1.1 and 0.33; for
(c), 1.0 and 0.35.\label{f:molecule}}
\end{center}
\end{figure}

Our results show that a two-component Fermi mixture beyond the critical
point can be effectively described as a strongly interacting BF mixture.
In Fig.~\ref{f:molecule}, we compare our experimental data with
numerically obtained density profiles~\cite{note2} without any adjustable
parameter, showing excellent agreement.  After including the LHY
correction, small discrepancies are visible only at the highest boson
densities exceeding $k_{F,\rb}a=1$, where one would expect unitarity
corrections.  It is surprising that we didn't need any beyond-mean-field
corrections for the boson-fermion interaction. Such corrections have been
calculated for a system of point bosons and fermions~\cite{Saam69}.
However, including them into our fit function degraded the quality of the
fit. Recent QMC simulations have shown that the equation of state of a
polarized Fermi gas on the BEC side is remarkably close to
$\cenergy=\cenergy_\mathrm{LHY}+\cenergy_{\rbf,\mathrm{M}}+\cenergy_\rf$
with $a_\rbb=0.6a$ and $a_\rbf=1.18a$ down to $1/k_{F\up}a
>0.5$~\cite{PG08,PS08} in agreement with our findings.  It appears
that the beyond-mean-field term is offset by other corrections, possibly
due to the composite nature of the bosons.  Further studies of this rich
system could address beyond-mean-field terms, characterize the break-down
of the BF description close to the critical point, and look for finite
temperature effects.

\begin{figure}
\begin{center}
\includegraphics[width=2.8in]{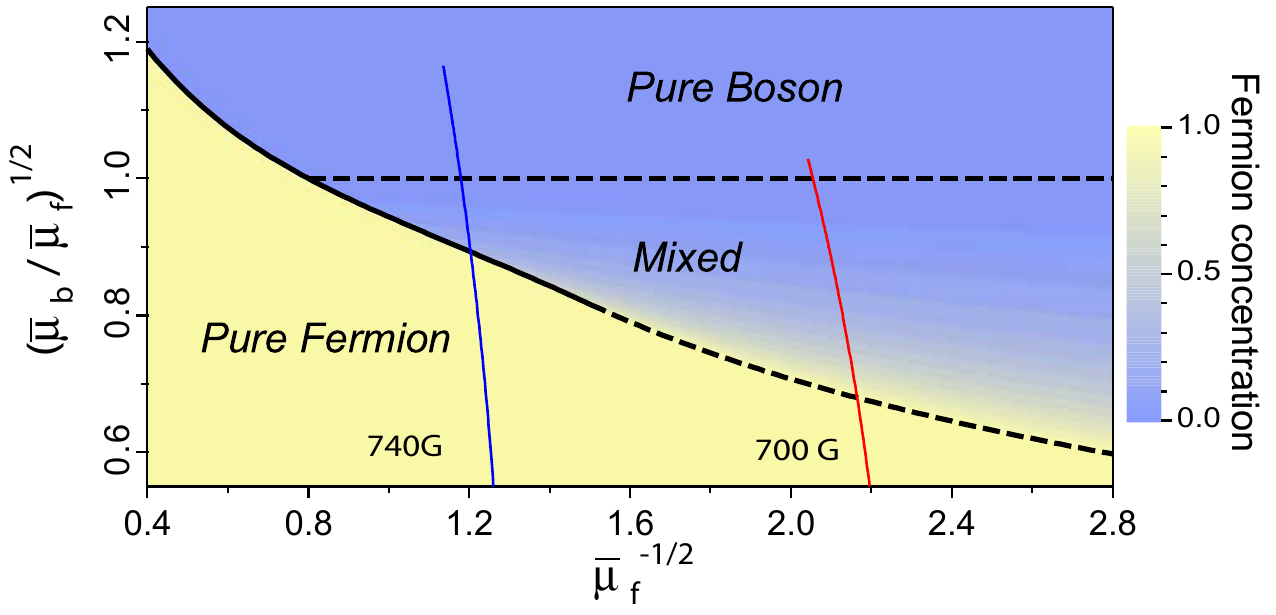}
\caption{(color online) Mean-field phase diagram of a dilute Bose-Fermi
mixture. $\bar{\mu}_\rf$ and $\bar{\mu}_\rb$ are the chemical potentials
for fermions and bosons in units of $\frac{\hbar^2}{2m_\rf} (6\pi^2
\bar{n}_\rf)^{2/3}$ and $\frac{4\pi \hbar^2 a_\rbb}{m_\rb} \bar{n}_\rb$,
respectively, where $\bar{n}_\rf= \frac{9\pi}{2}
\frac{a_\rbb^3}{a_\rbf^6} \frac{ m_\rb^3 m_\rf^3}{(m_\rb+m_\rf)^6}$ and
$\bar{n}_\rb=\frac{9\pi}{4} \frac{a_\rbb^2}{a_\rbf^5} \frac{ m_\rb^3
m_\rf^2}{(m_\rb+m_\rf)^5}$ (see Ref.~\cite{VPS00}). The thick lines
indicate the phase transitions (solid: first-order, dashed:
second-order). The thin lines represent typical cuts through the phase
diagram realized in our trapped samples.\label{f:BFphasediagram}}
\end{center}
\end{figure}

One motivation for the realization of BF mixtures is to extend studies of
$^3$He-$^4$He mixtures. With tunable interactions near Feshbach
resonances, cold atom systems can access a wider regime of the phase
diagram. Predicted phenomena include phase separation and
miscibility~\cite{molmer98,VPS00}, boson-mediated, effective
fermion-fermion coupling~\cite{BBP67,BHS00} and novel collective
excitations~\cite{Yip01,SGT04}. The density profiles in
Fig.~\ref{f:molecule} show a sharper boundary for higher magnetic fields.
This is consistent with Fig.~\ref{f:BFphasediagram} which predicts that
in the same magnetic field range, the transition from full miscibility to
phase separation has taken place.

An interacting BF system has been also realized in $^{87}$Rb-$^{40}$K
mixtures~\cite{OOH06,ZDF06}.  The \li system studied here has the
advantages of using a single atomic species and much longer lifetimes of
several seconds, but cannot access attractive boson-fermion interactions.

In conclusion, a two-component Fermi gas with population imbalance is a
realization of a long-lived strongly interacting BF mixture. This is a
new BF system with tunable interactions. Furthermore, it offers
intriguing possibilities to study the emergence of bosonic behavior from
a mixture of fermions.

We thank A. Keshet for a critical reading of the manuscript. This work was
supported by NSF, ONR, MURI and ARO Award W911NF-07-1-0493 (DARPA OLE
Program).


\begin{thebibliography}{23}
\bibitem{note1}
$\kappa$ is almost independent of the total population imbalance near the
resonance~\cite{Chevy06,Shin08}.
\bibitem{note2}
For a given $\cenergy$, the density profiles were numerically obtained to
be self-consistent with Eqs.~(\ref{e:muf}) and (\ref{e:mub}).
\end{thebibliography}
\end{document}